\documentclass[11pt,twoside]{article}

\usepackage{asp2006}
\usepackage{epsf}
\usepackage{graphicx}

\begin{document}
\title{Three Decades of Very Long Baseline Interferometry Monitoring of the Parsec-Scale Jet in 3C~345}   
\author{Frank K. Schinzel, Andrei P. Lobanov, and J.~Anton Zensus}   
\affil{Max-Planck-Institut f\"ur Radioastronomie,\\Auf dem H\"ugel 69, 53121 Bonn, Germany}    

\begin{abstract} 
The 16$^\mathrm{th}$ magnitude quasar 3C~345 (redshift z$=$0.5928) shows structural
and emission variability on parsec scales around a compact unresolved
radio core. For the last three
decades it has been closely monitored with very long baseline 
interferometry (VLBI), yielding a wealth of information about 
the physics of relativistic
outflows and dynamics of the central regions in AGN. We present here
preliminary results for the long-term jet evolution, based on the
15~GHz monitoring data collected by the MOJAVE survey and various
other groups over the last $\sim$14 years and combined with data from earlier
VLBI observations of 3C~345 which started in 1979. We discuss the
trajectories, kinematics, and flux density evolution of enhanced
emission regions embedded in the jet and present evidence for
geometrical (e.g. precession) and physical (e.g. relativistic shocks
and plasma instability) factors determining the morphology and
dynamics of relativistic flows on parsec scales.
\end{abstract}

\section{The Source}

The 16$^\mathrm{th}$ magnitude quasar 3C~345 (redshift z=0.5928) has
been observed at radio wavelengths for over 30 years, in particular with VLBI (cf.,
\citet{1986ApJ...308...93B}, \citet{1992A&A...257...31B},
\citet{1995ApJ...443...35Z}, \citet{1997ApJ...480..596U},
\citet{1999ApJ...521..509L}, \citet{2000A&A...354...55R},
\citet{2005A&A...431..831L}).
The source still continues to be of special interest due to its
complex, helical parsec-scale jet around a compact unresolved radio
core and its pronounced multiwavelength variability. A likely 8-10~years
periodicity of the high activity phases in 3C~345 has been identified
\citep{1999ApJ...521..509L}. Measurements of nuclear opacity and
magnetic field strength \citep{1998A&A...330...79L} yield a total mass
for the central engine of (4.0~$\pm$~2.4)~$\cdot\,10^9$~M$_\odot$. We
have analyzed VLBI observations of the last three decades in order to
understand the physics of the relativistic outflow and dynamics of
central regions in 3C~345. Preliminary results of this analysis are
presented here, focusing specifically on trajectories, kinematics,
and flux density evolution of enhanced emission regions embedded in
the jet.

\section{Observations}

We made use of a total of 201 observations (see Table~\ref{tab:obs})
that included Very Long Baseline Array (VLBA) observations obtained
from the NRAO Archive, observations by the MOJAVE survey, and
published values pre-dating the VLBA (before
1995)\footnote{\citet{1983ApJ...271..536U},
\citet{1992ApJ...398...74U}, \citet{1986ApJ...308...93B},
\citet{1995ApJ...443...35Z}, \citet{1992A&A...257...31B},
\citet{1996PhD...L}, \citet{1993A&A...275..375K},
\citet{2000A&A...354...55R}, \citet{2004PhD...K}}. Archival VLBI
observations were calibrated using NRAO's Astronomical Imaging
Processing System (AIPS). The total intensity and polarization data
was processed, with corrections applied for atmospheric opacity
(if deemed necessary), Faraday rotation, and Earth
orientation parameters used by the VLBA correlator. Fringe fitting was used to calibrate the
observations for group and phase delays. The then calibrated
visibility data was imaged using Caltech's Difmap
\citep{1995BAAS...27..903S}. The source structure was 
modelfitted using circular Gaussian components. At the end, individual Gaussian components
were cross-identified at different epochs in order
to follow the evolution of individual bright features in the jet.

It should be noted that the physical nature of these features, also
referred to as jet components, is still a matter of debate. Presently,
the common viewpoint is that moving jet features are relativistic
shocks in the jet plasma emitting optically thin synchrotron
radiation.

\begin{table}[!ht]
  \caption{Overview of VLBI observations used in this work.}
  \label{tab:obs}
  \smallskip
  \begin{center}
  {\small
  \begin{tabular}{cccc}
    \tableline\tableline
    \noalign{\smallskip}
    Time & \# of epochs & Frequencies & Type \\
    \noalign{\smallskip}
    \tableline
    \noalign{\smallskip}
    2009-     & 6+6+6 & 15,24,43~GHz & BS193, BS194 (VLBA) \\
    2002-2009 & 12 & 15~GHz & MOJAVE Survey (VLBA) \\
    1995-2009 & 61+7+7 & 15,22,43~GHz & archival re-reduced (VLBA) \\
    1979-1995 & 48+48 & 1-15,22-100~GHz & published data$^1$\\
    \noalign{\smallskip}
    \tableline
    \noalign{\smallskip}
    1979-2009 & Total: 201 \\
    \noalign{\smallskip}
    \tableline\tableline
  \end{tabular}
  }
  \end{center}
\end{table}

\section{Results}

\subsection{Component C9}

\begin{figure}[!ht]
	\centering
	\includegraphics[width=6.5cm]{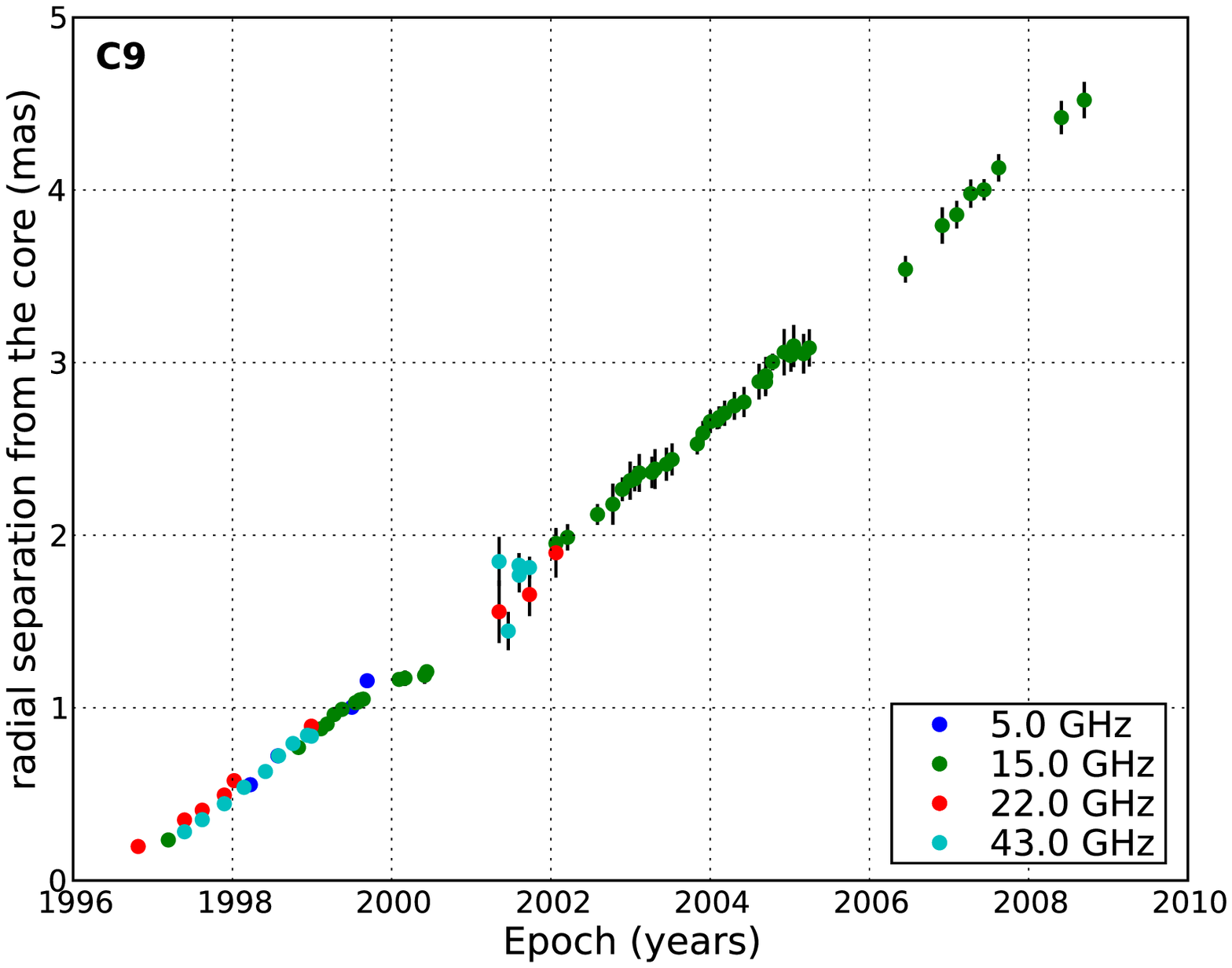}
	\includegraphics[width=6.5cm]{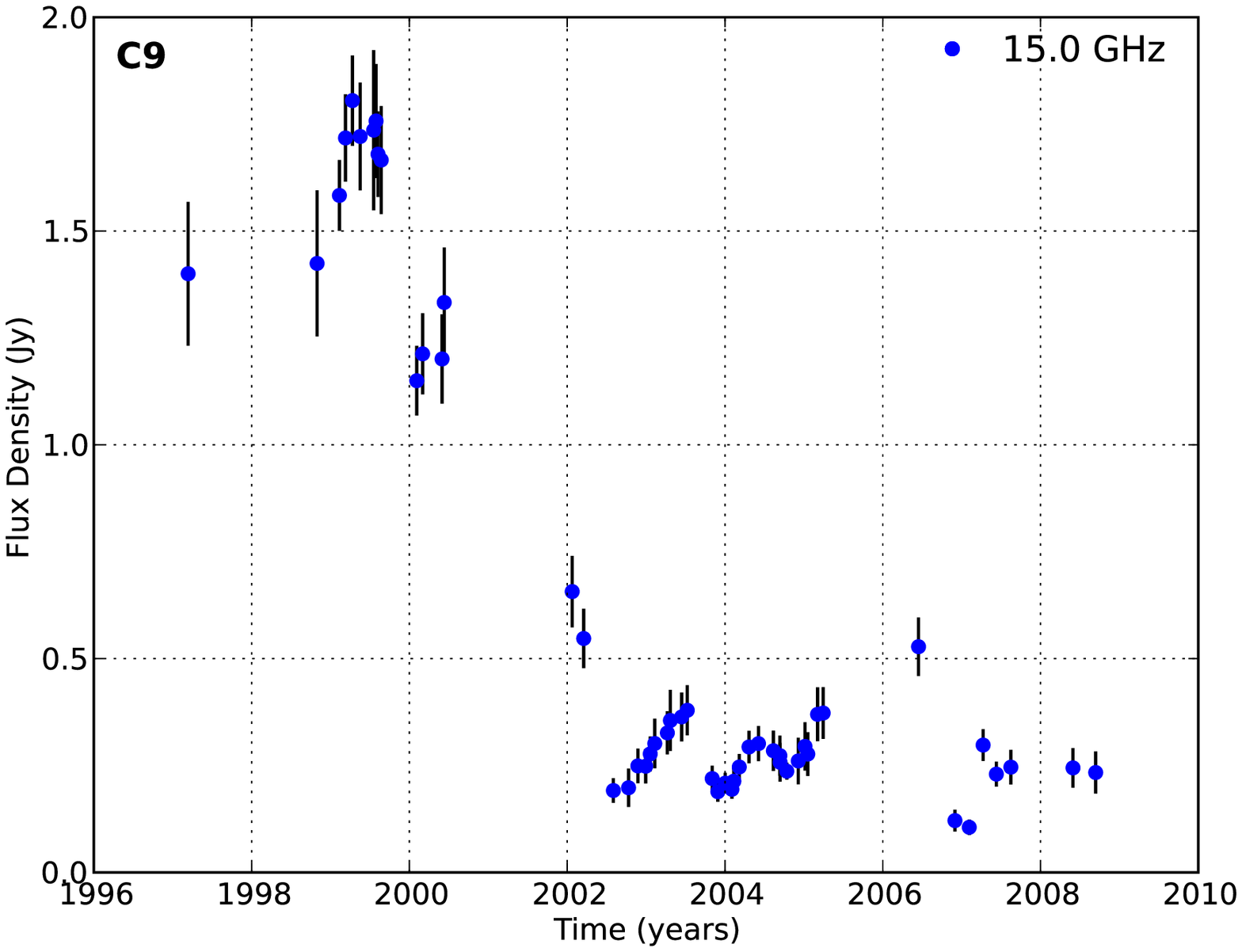}
	\caption{\textit{Left:} Radial separation from the core plotted over time for the individual jet feature labeled C9. The gap between 2000 and 2002 is due to difficulties of component identifications that will be resolved in the future. \textit{Right:} Flux density evolution of the jet feature C9 at 15~GHz for a period of 13 years.}
	\label{fig:C9-radialsep}
\end{figure}

The VLBI data collected on the pc-scale jet in 3C~345 reveals 16
bright features (labeled C1-C16, with C1 being the oldest feature)
that we are able to represent by circular Gaussian modelfits. Fits representing C1 were ignored in this analysis due to a lack of observations at $\geq$5~GHz as well as fits for C16, which appeared after 2007, were ignored.
The left plot of Figure~\ref{fig:C9-radialsep} shows the evolution of radial separation
of one of the jet components (C9) from the stationary core of the
jet. For observations other than at 15~GHz, a mean core-shift
referenced to the 15~GHz VLBI core position (defining the core-shift at 15~GHz to be 0~mas) has been determined and
applied, yielding an overall frequency-dependent position correction
of $\Delta r\left(\nu \right) = (1.37\cdot\nu (\mathrm{GHz})^{-1} - 0.058)$~mas. The
notable gap between 2000 and 2002 is due to an emerging new jet
feature causing substantial blending and making component
identification problematic. The data from that period are ignored at
the moment. The gaps 2005~-~2006 and 2007~-~2008 are due to a lack of
usable VLBI observations in these periods. The plot shows
a possible acceleration phase before 1998 and a subsequent transition
to a constant apparent speed. A linear fit to the observations around 1997 yields a time
of zero separation from the core of 1995.95~$\pm$~0.29. The proper
radial motion for a distance of $r<$~0.3~mas from the core is
0.099~mas~year$^{-1}$ and for a distance $r>$~0.3~mas of
0.378~mas~year$^{-1}$.

The 15~GHz flux density evolution of the component has a peak 
of (1.81 $\pm$~0.11)~Jy (Fig.~\ref{fig:C9-radialsep}:\textit{right}). The time of the peak is
3.43~$\pm$~0.31~years past the point of ejection. Analysis of the flux
density evolution of all components at 15~GHz gives an average time
of the peak after zero separation of $\sim$3.2~years.

\subsection{Trajectories}

\begin{figure}[!ht]
  \centering
  \includegraphics[width=8cm]{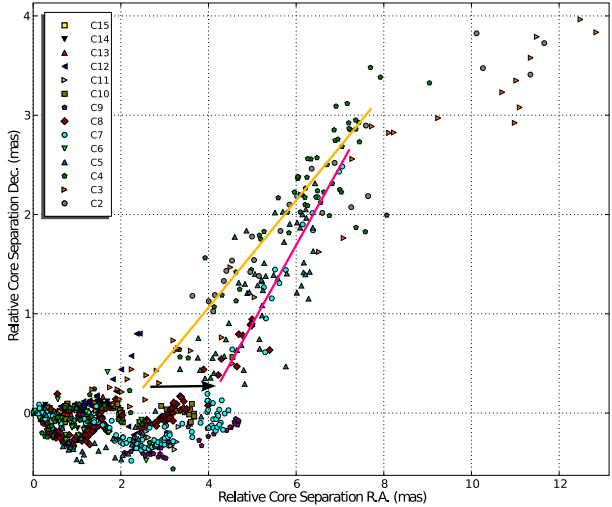}
  \caption{Trajectories of all jet features plotted on top of each other.}
  \label{fig:trajectories_all}
\end{figure}

Figure~\ref{fig:trajectories_all} shows the trajectories of all jet
features. The jet is traced up to about 15\,mas distance from
the core. The measurements become more sparse at distances r$>$8~mas
as a result of less frequent observations at frequencies $\leq$8~GHz.

The jet is initially directed westward, at a position angle of almost
90$\deg$. The individual jet features trace a common channel of
$\sim$1~mas in width, within a distance of $\sim$5~mas from the
core. Beyond this distance the jet sharply turns northwards. 

This northward turning evolves in time. Earlier components were
turning at around 3~mas, more recent features turn at 5~mas. Evidence
for long-term changes are seen at shorter distances as well.

\begin{figure}
  \centering
  \includegraphics[width=8cm]{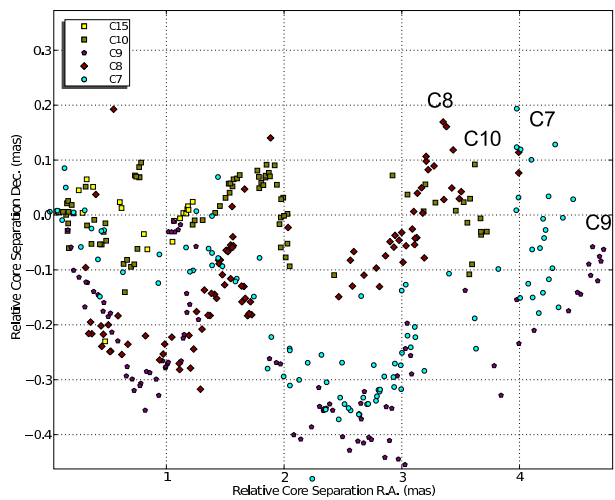}
  \caption{Close-up view of the trajectories for separations $<$5~mas.}
  \label{fig:trajectories_small}
\end{figure}

In Figure~\ref{fig:trajectories_small}, the trajectories in the
region up to 5~mas separation from the VLBI core are shown. It looks
like subsequent features follow slightly different paths. The
northward turning points evolve in a way that C8 follows C10, C7 and
then C9. It looks like a wiggling of the jet turning point on smaller
scales. A behavior like this is expected for a helical, precessing
jet. This needs to be testedquantitativelyy in future work. The maximum relative 
core separation in declination varies from component to component. This is
especially evident in the region up to 1.5~mas from the jet. C7 shows
a maximum separation of $\sim$0.1~mas, but C8 gets up to $\sim$0.2~mas. followed by C9 at
$\sim$0.3~mas. After this, C10 as well as C11 come back to
$\sim$0.05~mas. This behavior is curious and a precessing jet model
will be tried in order to explain this as well.

\subsection{Evolution of the component ejection angle}

\begin{figure}
  \centering
  \includegraphics[width=8cm]{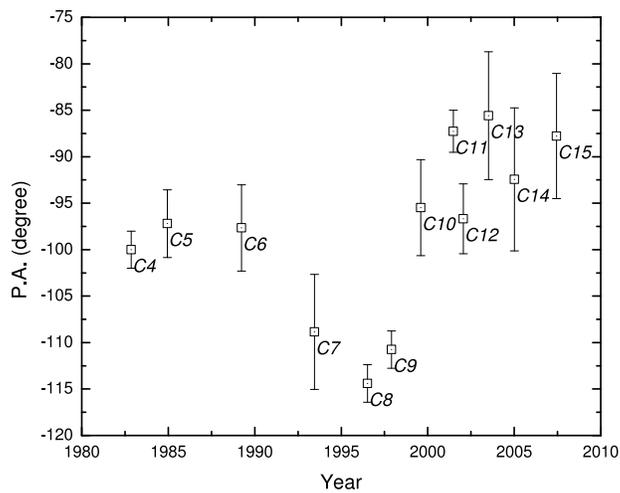}
  \caption{Plot of position angles for each jet feature at a separation of 0.5~mas from the core vs time.}
  \label{fig:pa-evolution}
\end{figure}

The component position angles measured at 0.5~mas radial distance from
the VLBI core at 15~GHz offer a different way to represent the previously
described deviations from the 90$\deg$ position angle (core separation in declination) 
in the trajectories of subsequent jet features C4-C15
(1983-2007), as shown in Figure~\ref{fig:pa-evolution}. We see no
clear periodic trends as has been claimed in the
past. \citet{2005A&A...431..831L} and \citet{2003enig.conf...92K}
describe in their work a short-term periodicity of 8-10 years with an
underlying long-term trend of 0.4-2.6$\deg$~year$^{-1}$. We cannot
confirm this here, but we see long-term variability as well as short
term changes in the position angles since 2000. The reason these short
term changes are not seen in earlier observations is due to the lower
sensitivity and sampling of observations pre-dating the VLBA (before
1995). At that time only the brightest features were detectable. With
recent, more sensitive and more frequent observations, we are able to
see much more structure in the jet and are even able to see the bright
edges of the jet. As stated above, we need to check whether the
observed behavior can be brought into agreement with a
precessing-helical jet model.

\subsection{Apparent velocities}

Apparent velocities $\beta_\mathrm{app}$ have been determined for all
components. A proper motion of 1 mas year$^{-1}$ is translated to a
$\beta_\mathrm{app}$ of 19.7 h$^{-1}$c in concordance with the
standard $\Lambda$CDM (H$_0$ = 70 km s$^{-1}$ Mpc$^{-1}$ and
$\Omega_\Lambda$ = 0.72 = 1 - $\Omega_\mathrm{M}$, h is the dimensionless Hubble parameter).  We have not been able to
determine a common velocity for separations $<$0.7~mas; however we
find an upper limit of $\beta_\mathrm{app}\leq$2.25 h$^{-1}$c. For a
separation $>$0.7~mas, we obtained a $\beta_\mathrm{app}\approx$7.0
h$^{-1}$c for most of the components, with the exception of two
features that show an apparent velocity of $\approx$9 h$^{-1}$c.

\section{Summary \& Outlook}

3C~345 shows a complex jet morphology with many bright features
observed by VLBI which have been traced for three decades now. We
found a consistent superluminal motion of moving jet features at
distances $>$0.7~mas and an apparent acceleration from smaller
velocities at closer distances. We saw evidence for long- and
short-term evolution in the jet trajectories of subsequent VLBI
features as expected by a precessing jet with helical morphology.

We are going to expand this observational analysis to study spectral
evolution as well as coreshifts. We are going to test various models
in order to explain the underlying physics of the observed phenomena
quantitatively.

\acknowledgements 
\begin{small}Frank Schinzel was supported for this research through a stipend from
the International Max Planck Research School (IMPRS) for
Astronomy and Astrophysics at the Universities of Bonn
and Cologne. The National Radio Astronomy Observatory is a facility of the National Science Foundation operated under cooperative agreement by Associated Universities, Inc. This research has made use of data from the MOJAVE database that is maintained by the MOJAVE team \citep{2009AJ....137.3718L}.\end{small}



\begin{thebibliography}{}
	\bibitem[Baath et al.(1992)]{1992A&A...257...31B} Baath, L.~B., et al.\ 1992, \aap, 257, 31 
	\bibitem[Biretta et al.(1986)]{1986ApJ...308...93B} Biretta, J.~A., Moore, R.~L., \& Cohen, M.~H.\ 1986, \apj, 308, 93 
	\bibitem[Klare et al.(2003)]{2003enig.conf...92K} Klare, J., Zensus, J.~A., Witzel, A., Krichbaum, T.~P., Lobanov, A.~P., \& Ros, E.\ 2003, Proceedings of the Second ENIGMA Meeting, 92 
	\bibitem[Klare (2004)]{2004PhD...K} Klare, J., PhD Thesis, Max-Planck Institut f\"ur Radioastronomie/Rheinische Friedrich-Wilhelms-Universit\"at Bonn, Bonn, 2004 
	\bibitem[Krichbaum et al.(1993)]{1993A&A...275..375K} Krichbaum, T.~P., et al.\ 1993, \aap, 275, 375 
	\bibitem[Lister et al.(2009)]{2009AJ....137.3718L} Lister, M.~L., et al.\ 2009, \aj, 137, 3718 
	\bibitem[Lobanov (1996)]{1996PhD...L} Lobanov, A.~P., PhD Thesis, New Mexico Institute for Mining \& Technology, Socorro, New Mexico, 1996
	\bibitem[Lobanov(1998)]{1998A&A...330...79L} Lobanov, A.~P.\ 1998, \aap, 330, 79 
	\bibitem[Lobanov \& Zensus(1999)]{1999ApJ...521..509L} Lobanov, A.~P., \& Zensus, J.~A.\ 1999, \apj, 521, 509 
	\bibitem[Lobanov \& Roland(2005)]{2005A&A...431..831L} Lobanov, A.~P., \& Roland, J.\ 2005, \aap, 431, 831 
	\bibitem[Ros et al.(2000)]{2000A&A...354...55R} Ros, E., Zensus, J.~A., \& Lobanov, A.~P.\ 2000, \aap, 354, 55 
	\bibitem[Shepherd et al.(1995)]{1995BAAS...27..903S} Shepherd, M.~C., Pearson, T.~J., \& Taylor, G.~B.\ 1995, \baas, 27, 903 
	\bibitem[Unwin et al.(1983)]{1983ApJ...271..536U} Unwin, S.~C., Cohen, M.~H., Pearson, T.~J., Seielstad, G.~A., Simon, R.~S., Linfield, R.~P., \& Walker, R.~C.\ 1983, \apj, 271, 536 
	\bibitem[Unwin \& Wehrle(1992)]{1992ApJ...398...74U} Unwin, S.~C., \& Wehrle, A.~E.\ 1992, \apj, 398, 74 
	\bibitem[Unwin et al.(1997)]{1997ApJ...480..596U} Unwin, S.~C., Wehrle, A.~E., Lobanov, A.~P., Zensus, J.~A., Madejski, G.~M., Aller, M.~F., \& Aller, H.~D.\ 1997, \apj, 480, 596 
	\bibitem[Zensus et al.(1995)]{1995ApJ...443...35Z} Zensus, J.~A., Cohen, M.~H., \& Unwin, S.~C.\ 1995, \apj, 443, 35 
\end{thebibliography}
\end{document}